\documentclass{article}
\usepackage{graphicx}
\usepackage{amsmath}

\input{tcilatex}

\begin{document}

\begin{center}
{\Large GRAVITY CANNOT BE QUANTIZED}
\end{center}

{\Large \bigskip }

\begin{center}
\noindent M. E. DE SOUZA

\noindent \textit{Department of Physics, Universidade Federal de Sergipe, }

\textit{Campus Universit\'{a}rio, 49100-000 S\~{a}o Crist\'{o}v\~{a}o, SE,
Brazil}
\end{center}

\bigskip

\parbox{4.3in}
{\noindent {\small Abstract. - Taking a deeper look at the fundamental force of
gravity one arrives at the conclusion that it is quite an unusual field 
because it does not have a fermion associated to it. And the absence of such
fermion shadows the existence of the graviton itself. Therefore, gravity
quantization is also doubtful.}

{\small \medskip }\noindent 

\noindent \textit{Keyworks: }gravity quantization, graviton, 

\medskip

\noindent PCAS Nos.: 96.40.Tv; 13.15.+g; 14.60.Lm }

\bigskip

Any elementary particle is either a boson or a fermion. Bosons are mediators
of the interactions and fermions are carriers of the quantized charges. As
an example in the electromagnetic interaction the electron current is given
by

\begin{equation}
j_{V}^{\mu }=e\overline{\psi }\gamma ^{\mu }\psi
\end{equation}
where $e$ is the quantum of charge, and $\psi $ is the fermion Dirac spinor
of the electron. In the case of the pionic current in the domain of nuclear
physics one has 
\begin{equation}
j_{S,\pi N}=g_{\pi N}\overline{\psi }\gamma ^{5}(\mathbf{\tau .\Phi })\psi
\end{equation}
in which $g_{\pi N}$ is the strong charge, $\mathbf{\tau }$ are the isospin
Pauli matrices, $\psi $ is the nucleonic isospinor 
\begin{equation}
\psi =\left( 
\begin{array}{c}
\varphi _{p} \\ 
\varphi _{n}
\end{array}
\right) .
\end{equation}

\noindent and $\mathbf{\Phi }$ is the isovector 
\begin{equation}
\mathbf{\Phi }=\left( 
\begin{array}{c}
\phi _{\pi ^{+}} \\ 
\phi _{\pi ^{0}} \\ 
\phi _{\pi ^{-}}
\end{array}
\right)
\end{equation}
where $\phi $ are pseudoscalar functions.

Following the reasoning above developed it is very important to ask if
gravity has a charge carrier. If it has one it is not known yet. And it may
not exist after all. A is well known the gravitational field is rather odd.
General relativity has shown either experimentally or theoretically that
massless particles are attracted gravitationally by massive particles.
According to general relativity the 4-momentum of a freely moving photon is
written as$^{1}$%
\begin{equation}
\nabla _{p}p=0
\end{equation}
where the four-momentum of the photon is $p=\frac{d}{d\lambda }$ and $%
\lambda $ is an affine parameter. This geodesic equation can be written as 
\begin{equation}
\frac{dp^{\alpha }}{d\lambda ^{\ast }}+\Gamma _{\beta \gamma }^{\alpha
}p^{\beta }p^{\gamma }=0
\end{equation}
from which it can be shown that a photon(\emph{zero mass}) suffers a
deflection given by the angle 
\begin{equation}
\Delta \phi =4M/b=1^{\prime \prime }.75(R_{\odot }/b)
\end{equation}
in which $M$ is the sun's mass, $R_{\odot \text{ }}$is the sun's radius and $%
b$ is the impact parameter. This means that even particles with
gravitational \ \ ``charge '' equal to zero suffer the influence of gravity.
On the other hand in Newtonian gravity the gravitational potential energy
between two massive bodies is 
\begin{equation}
E_{p}(r)=-G\frac{m_{1}m_{2}}{r}
\end{equation}
which is of Yukawa type. According to this equation the two masses are the
two gravitational charges. And if the fermionic mass carrier exists each
mass is a multiple of the fermion mass. Otherwise, mass cannot be quantized
because without a fermionic mass carrier there cannot be mass currents. How
can gravity be quantized without quantizing mass and without fermionic
currents? \ 

The quantization of gravity has to exist either in curved spacetime or in
flat spacetime. For example, it is expected that \textbf{if a body is
excited gravitationally it should emit gravitational charge carriers into
space and it is also expected that when a body is gravitationally excited
the charge carriers (fermions) should change quantum states. When particles
change mass in a high energ y collision there should exist such fermion
currents.}

Let us admit the existence of such mass carrier and let us call it \emph{%
masson}. Since it is a fermion, in flat space it has to satisfy Dirac
equation which written in covariant form is (a free fermion) 
\begin{equation}
(i\hbar \gamma ^{\mu }\partial _{\mu }-mc)\psi =0.
\end{equation}
In the case of a vectorial mass current one has 
\begin{equation}
j^{\mu }=cm\overline{\psi }\gamma ^{\mu }\psi
\end{equation}

\noindent because it is expected that its mass is also its charge. From
Dirac equation we have $i\hbar \gamma ^{\nu }\partial _{\nu }\psi =cm\psi $
and 
\begin{equation*}
i\hbar \overline{\psi }\gamma ^{\mu }\gamma ^{\nu }\partial _{\nu }\psi =cm%
\overline{\psi }\gamma ^{\mu }\psi =j^{\mu }
\end{equation*}
As $\gamma _{\mu }\gamma ^{\mu }=4$, we can write 
\begin{equation}
i\hbar \overline{\psi }\gamma ^{\mu }\gamma ^{\nu }\partial _{\nu }(\gamma
_{\mu }\gamma ^{\mu }\psi )=i\hbar \overline{\psi }\gamma ^{\mu }\gamma
^{\nu }\gamma _{\mu }\partial _{\nu }(\gamma ^{\mu }\psi )=4mc\overline{\psi 
}\gamma ^{\mu }\psi .
\end{equation}
Since $\gamma ^{\mu }\psi $ is also a solution of Dirac equation we obtain 
\begin{equation}
i\hbar \overline{\psi }\gamma ^{\nu }\partial _{\nu }(\gamma ^{\mu }\psi )=mc%
\overline{\psi }\gamma ^{\mu }\psi
\end{equation}
and 
\begin{equation}
i\hbar \overline{\psi }\gamma ^{\nu }\gamma ^{\mu }\gamma _{\mu }\partial
_{\nu }(\gamma ^{\mu }\psi )=4mc\overline{\psi }\gamma ^{\mu }\psi
\end{equation}
Since $\gamma ^{\nu }\gamma ^{\mu }+\gamma ^{\mu }\gamma ^{\nu }=2g^{\nu
}{}^{\mu }$, summing up equations (11) and (13) we obtain

\begin{equation}
i\hbar \overline{\psi }g^{\nu }{}^{\mu }\gamma _{\mu }\partial _{\nu }\gamma
^{\mu }\psi =4mc\overline{\psi }\gamma ^{\mu }\psi
\end{equation}

\noindent where $g^{\nu }{}^{\mu }$ is the metric 
\begin{equation}
g^{\nu \mu }{}=\left( 
\begin{array}{cccc}
1 & 0 & 0 & 0 \\ 
0 & -1 & 0 & 0 \\ 
0 & 0 & -1 & 0 \\ 
0 & 0 & 0 & -1
\end{array}
\right) .
\end{equation}

\noindent Therefore, we obtain the fermionic mass operator (of the \emph{%
masson}) 
\begin{equation}
m=\frac{i}{4c}\hbar g^{\nu }{}^{\mu }\gamma _{\mu }\partial _{\nu }
\end{equation}
and the mass current 
\begin{equation}
j^{\mu }=\frac{i\hbar }{4}\overline{\psi }g^{\nu }{}^{\mu }\gamma _{\mu
}\partial _{\nu }\gamma ^{\mu }\psi =\frac{i\hbar }{4}\overline{\psi }g^{\mu
\nu }{}\partial _{\nu }\gamma _{\mu }\gamma ^{\mu }\psi =i\hbar \overline{%
\psi }g^{\mu \nu }{}\partial _{\nu }\psi .
\end{equation}
These two equations clearly show that the \emph{masson }mass depends on the
metric. In curved space-time we can always choose a small region where
space-time is approximately flat. Hence, we can extend the meaning of $%
g^{\nu }{}^{\mu }$ to include curved space-time. Doing this we notice that
since the \emph{masson } mass depends on the metric it can not be unique,
that is, it has different values in different curved space-times. Since flat
space time is a local approximation of curved space-time its mass has only a
local meaning. Therefore, we stumbled into another obstacle in quantizing
gravity. We can do this formally. Let us take an orthogonal metric, that is
a metric in which $g^{\nu }{}^{\mu }=0$, for $\nu \neq \mu $. We have then
the metric 
\begin{equation}
g^{\nu \mu }{}=\left( 
\begin{array}{cccc}
g_{00} & 0 & 0 & 0 \\ 
0 & g_{11} & 0 & 0 \\ 
0 & 0 & g_{22} & 0 \\ 
0 & 0 & 0 & g_{33}
\end{array}
\right) .
\end{equation}
If we are in a very small region of curved space-time (without large
curvature) we can say that $g_{00}\approx 1+f_{00}$, $g_{11}\approx
-1+f_{11} $, $g_{22}\approx -1+f_{22}$ and $g_{33}\approx -1+f_{33}$, and we
have for small $f_{jj}$ 
\begin{equation}
\delta m=\frac{i}{4c}\hbar \Delta ^{\nu }{}^{\mu }\gamma _{\mu }\partial
_{\nu }
\end{equation}
with 
\begin{equation*}
\Delta ^{\nu }{}^{\mu }=\left( 
\begin{array}{cccc}
f_{00} & 0 & 0 & 0 \\ 
0 & f_{11} & 0 & 0 \\ 
0 & 0 & f_{22} & 0 \\ 
0 & 0 & 0 & f_{33}
\end{array}
\right)
\end{equation*}

\noindent where $\delta m$ is non-Euclidean. This is the mass acquired by
the masson directly from curvature.

If a scalar mass current $j=cm\overline{\psi }\psi $\ is used one obtains a
similar result as is proven as follows. From Dirac equation 
\begin{equation}
i\hbar \gamma ^{\nu }\partial _{\nu }\psi =cm\psi
\end{equation}
\noindent\ and since $\gamma ^{\mu }\psi $ is also a solution 
\begin{equation}
i\hbar \gamma ^{\nu }\partial _{\nu }\gamma ^{\mu }\psi =cm\gamma ^{\mu
}\psi .
\end{equation}
Multiplying this equation from the left by $\gamma _{\mu }$ and taking into
account that $\gamma _{\mu }\gamma ^{\mu }=4$\ one obtains 
\begin{equation}
i\hbar \gamma _{\mu }\gamma ^{\nu }\gamma ^{\mu }\partial _{\nu }\psi
=mc\gamma _{\mu }\gamma ^{\mu }\psi =4mc\psi
\end{equation}

\noindent But multiplying Eq. 20 from the left by $\gamma _{\mu }\gamma
^{\mu }(=4)$ one has 
\begin{equation}
i\hbar \gamma _{\mu }\gamma ^{\mu }\gamma ^{\nu }\partial _{\nu }\psi
=4mc\psi
\end{equation}
and upon summation of these two last equations 
\begin{equation}
i\hbar \gamma _{\mu }(\gamma ^{\nu }\gamma ^{\mu }+\gamma ^{\mu }\gamma
^{\nu })\partial _{\nu }\psi =8mc\psi
\end{equation}
and as $\gamma ^{\nu }\gamma ^{\mu }+\gamma ^{\mu }\gamma ^{\nu }=2g^{\nu
\mu }$ one gets 
\begin{equation}
i\hbar \gamma _{\mu }g^{\nu \mu }\partial _{\nu }\psi =4cm\psi
\end{equation}
and the mass operator 
\begin{equation}
m=\frac{i}{4c}\hbar \gamma _{\mu }g^{\nu }{}^{\mu }\partial _{\nu }
\end{equation}
is obtained and the mass current 
\begin{equation*}
j=cm\overline{\psi }\psi =\overline{\psi }\frac{i}{4c}\hbar \gamma _{\mu
}g^{\nu }{}^{\mu }\partial _{\nu }\psi
\end{equation*}
and of course the same discussion above, done in the case of a vector
current, continues to be valid.

If the current is a pseudoscalar current then

\begin{equation}
j=cm\overline{\psi }\gamma ^{5}\psi .
\end{equation}
From Dirac equation 
\begin{equation}
i\hbar \gamma ^{\nu }\partial _{\nu }\psi =cm\psi
\end{equation}
which multiplied by $\gamma ^{5}$ is 
\begin{equation}
i\hbar \gamma ^{5}\gamma ^{\nu }\partial _{\nu }\psi =cm\gamma ^{5}\psi .
\end{equation}
When this eqation is multiplied by $\gamma _{\mu }\gamma ^{\mu }(=4)$ it
becomes 
\begin{equation}
i\hbar \gamma _{\mu }\gamma ^{\mu }\gamma ^{5}\gamma ^{\nu }\partial _{\nu
}\psi =4cm\gamma ^{5}\psi
\end{equation}
which is equal to 
\begin{equation}
-i\hbar \gamma _{\mu }\gamma ^{5}\gamma ^{\mu }\gamma ^{\nu }\partial _{\nu
}\psi =4cm\gamma ^{5}\psi .
\end{equation}
And as $\gamma ^{\mu }\psi $ is also a solution of Dirac equation 
\begin{equation}
i\hbar \gamma ^{\nu }\partial _{\nu }\gamma ^{\mu }\psi =cm\gamma ^{\mu }\psi
\end{equation}
and since the left side is equal to $i\hbar \gamma ^{\nu }\gamma ^{\mu
}\partial _{\nu }\psi $ one has 
\begin{equation}
i\hbar \gamma ^{5}\gamma ^{\nu }\gamma ^{\mu }\partial _{\nu }\psi =cm\gamma
^{5}\gamma ^{\mu }\psi
\end{equation}
and 
\begin{equation}
i\hbar \gamma _{\mu }\gamma ^{5}\gamma ^{\nu }\gamma ^{\mu }\partial _{\nu
}\psi =cm\gamma _{\mu }\gamma ^{5}\gamma ^{\mu }\psi =-cm\gamma _{\mu
}\gamma ^{\mu }\gamma ^{5}\psi
\end{equation}
or 
\begin{equation}
-i\hbar \gamma _{\mu }\gamma ^{5}\gamma ^{\nu }\gamma ^{\mu }\partial _{\nu
}\psi =cm\gamma _{\mu }\gamma ^{\mu }\gamma ^{5}\psi =4cm\gamma ^{5}\psi .
\end{equation}
Summing up eqs. 31 and 35 the mass operator 
\begin{equation}
m\gamma ^{5}=-\frac{i\hbar }{2c}\gamma _{\mu }\gamma ^{5}g^{\nu \mu
}\partial _{\nu }
\end{equation}
is obtained which depends also on the metric.

If the same procedure is done for a pseudovectorial current 
\begin{equation}
j=cm\overline{\psi }\gamma ^{\mu }\gamma ^{5}\psi 
\end{equation}
one arrives at the mass operator \ 
\begin{equation}
m=\frac{i\hbar }{2c}g^{\nu \mu }\gamma ^{5}\gamma _{\mu }\partial _{\nu }.
\end{equation}
In the case of a tensorial antisymmetric current 
\begin{equation}
j=cm\overline{\psi }\sigma ^{\mu \nu }\psi 
\end{equation}
($\sigma ^{\mu \nu }=\frac{i}{2}(\gamma ^{\mu }\gamma ^{\nu }-\gamma ^{\nu
}\gamma ^{\mu }$) one easily finds that the \textbf{mass operator is
independent of the metric} and given by 
\begin{equation}
m=\frac{i\hbar }{c}\gamma ^{\nu }\partial _{\nu }.
\end{equation}

Since $1$, $\gamma ^{5}$, $\gamma ^{\mu }$, $\gamma ^{\mu }\gamma ^{5}$, and 
$\sigma ^{\mu \nu }$ form a basis for the space of all $4\times 4$ matrices,
any other tensor $\chi ^{\mu \nu }$ can be written in terms of a linear
combination of these 16 matrices and, therefore, symmetric mass currents
fall into the above categories already taken care of. Thus, the
gravitational field cannot be scalar, pseudoscalar, vectorial,
pseudovectorial, and symmetric tensorial field. The only possibility left is
to be an antisymmetric tensorial field which is a result that agrees well
with general relativity. Misner, Thorne and Wheeler$^{2}$ have proven that
the \textit{classical }gravitational field is an antisymmetric tensorial
field. This work shows that the same should hold quantum mechanically. But
since elementary fermions are spin $\frac{1}{2}$ particles we expect the 
\textit{masson} to be also an elementary fermion. But this means that the
graviton should have spin equal to $0$ or $1$ because bosons intermediate
states between fermions. In other words if the \textit{masson} is in a state
with spin $m_{s}=+\frac{\hbar }{2}$ it can only go to a state with $m_{s}=-%
\frac{\hbar }{2},+\frac{\hbar }{2}$ and this means that there should be the
emission of a graviton with spin equal to $1$ or $0$ but as was shown above
this possibility cannot happen. The existence of massons are thus
incompatible with the existence of the graviton. \textbf{Therefore, the
gravitational field can not be quantized and, of course, neither the masson
nor the graviton exists. This leads us to say that the gravitational field
is always a static field which is in line with the null results of
gravitational waves. }

\bigskip

\noindent {\Large References}

\bigskip

\noindent 1. C.W. Misner, K.S. Thorne and J.A.\ Wheeler, in Gravitation,
W.H. Freeman and Company, San Francisco, 1973, p. 446.

\noindent 2. Idem, pp 178-186.

\end{document}